\documentstyle[12pt]{article}
\input epsf.tex

\newcommand{\beq}{\begin{equation}}
\newcommand{\eeq}{\end{equation}}
\newcommand{\be}{\begin{equation}}
\newcommand{\ee}{\end{equation}}
\newcommand{\bea}{\begin{eqnarray}}
\newcommand{\eea}{\end{eqnarray}}

\def\href#1#2{#2}

\begin{document}

\baselineskip=15.5pt
\pagestyle{plain}
\setcounter{page}{1}

\begin{titlepage}
\begin{flushleft}
       \hfill                      {\tt hep-th/0806.xxxx}\\
       \hfill                       FIT HE - 08-02 \\
       \hfill                       KYUSHU-HET-111 \\
       \hfill                       Kagoshima HE - 08-1 \\
\end{flushleft}
\vspace*{3mm}

\begin{center}
{\huge Multi-quark baryons and color screening \\
\vspace*{2mm}
at finite temperature}
\end{center}
\vspace{5mm}

\begin{center}
\vspace*{5mm}
{\large Kazuo Ghoroku${}^{\dagger}$\footnote[1]{\tt gouroku@dontaku.fit.ac.jp},
Masafumi Ishihara${}^{\ddagger}$\footnote[2]{\tt masafumi@higgs.phys.kyushu-u.ac.jp},
${}^{\S}$Akihiro Nakamura\footnote[3]{\tt nakamura@sci.kagoshima-u.ac.jp}
${}^{\P}$Fumihiko Toyoda\footnote[4]{\tt ftoyoda@fuk.kindai.ac.jp}
%
}\\

\vspace*{2mm}

{${}^{\dagger}$Fukuoka Institute of Technology, Wajiro, 
Higashi-ku} \\
{
Fukuoka 811-0295, Japan\\}
{
${}^{\ddagger}$Department of Physics, Kyushu University, Hakozaki,
Higashi-ku}\\
{
Fukuoka 812-8581, Japan\\}
{
${}^{\S}$Department of Physics, Kagoshima University, Korimoto 1-21-35, \\Kagoshima 890-0065, Japan\\}
{
${}^{\P}$School of Humanity-Oriented Science and
Engineering, Kinki University,\\ Iizuka 820-8555, Japan}

\end{center}

\vspace*{8mm}
\begin{center}
{\large Abstract}
\end{center}
\noindent

We study baryons in $SU(N)$ gauge theories at finite temperature 
according to the gauge/string correspondence based on IIB string theory.  
The baryon is constructed out of D5 brane and $N$ fundamental strings
to form a color singlet $N$-quark bound state. 
At finite temperature and in the deconfining phase, 
we could find $k(<N)$-quark ``baryons''. Thermal properties of such $k$-quark baryons 
and also of the $N$-quark baryon are examined. We study
the temperature dependence of
color screening distance and Debye length of the baryon
of $k$-quark and $N$-quark. We also estimate the melting temperature, where
the baryons decay into quarks and gluons completely.

\vfill
\begin{flushleft}
\end{flushleft}
\end{titlepage}
\newpage

\section{Introduction}
In the context of string/gauge theory correspondence 
the baryons related to the D5-branes has been proposed \cite{wittenbaryon, 
groguri,imamura,cgs,baryonsugra,imafirst,cgst,ima-04,ima-05} and studied 
using the Born-Infeld approach \cite{calmal,gibbons,sin}. 
In these approaches, the fundamental strings
(F-string) are dissolved in the D5 brane in order to make
a baryon vertex. Here the D5-brane
is embedded as a probe in a 10D background in which
the corformal invariance is broken by the non-trivial dilaton. 
As a result of non-conformal invariance,
the embedded configurations of the D5-brane have in general cusps, which are the
singular points to be cancelled out by some external objects. This is
performed by introducing F-strings whose surface term cancells out with this
cusp singularity. This is represented by the no-force condition between
the F-strings and the D5-brane. From a viewpoint of smoothed picture of
the baryon configuration,
these added F-strings could be regarded as
the one which flow from the D5-brane through the cusps.

According to the above idea, an analysis
for the baryon configurations has been performed in \cite{GI} based on a simple
holographic model. 
At finite temperature and in the quark deconfinement
phase, we could find a possible bound states of small number $k(<N)$
of quarks with D5 vertex.
Here we study the thermal
properties of these states and also the usual $N$-quarks baryons.
The quarks in the baryon are
connected to the vertex through F-string. This F-string is common to
the one of mesons which are the bound states of a quark and an anti-quark.
So we expect, for the part related to the
quarks in the baryon, that the thermal properties are similar 
to the one of the mesons.
But the situation is different from the meson case in the 
following two points. (i) Since 
the baryon is constructed from $N$ quarks, various configurations are possible in the
quark deconfinement phase. They are discriminated by the number of quarks 
retained in the
state. (ii) The second point is the non-trivial
configuration of the vertex given by D5 brane, which is determined
by the dynamics of the gauge theory. 

These properties depend on the temperature (T). 
At low temperature, we find four types ((A) $\sim$ (D)) of D5-brane canfigurations.
Two of them ((C) and (D))
appear for the first time at finite temperature. When the temperature
exceeds a value
$T_{c_1}$, which is given in \cite{GI}, the solution is reduced to the one type
((A)).
But before arriving at this temperature, any baryon state decays into quarks
and gluons since its energy ($E$) exceeds over the sum of the effective mass 
of free quarks.
We call this temperature as melting point, $T_{\rm melt}$,
which is slightly smaller than
$T_{c_1}$ as estimated below.

The baryon energy $E$ varies with its size.
We study here the
relation between the size and the energy $E$ of the baryon.
As for the size of the baryon, we could measure it as the F-string length
observed in the real three dimensional space.
The configurations of the F-strings are intimately related
to the vertex configurations or the D5 brane structure, which varies with
the temperature. Through the study of the F-string configurations,
we find that the quarks in the baryon can not stay far from the vertex
exceeding a distance ($L_{\rm max}$) which increases with decreasing $T$.
This is the reflection of the thermal screening
of the color force and it 
is seen by measuring
$L_{q-v}$, the distance of the quarks
from the vertex, as a function of $E$. At low temperature,
we find $E-L_{q-v}$ relation as three connected curves. 
They are a little
complicated compared to the case of mesons. 
The first curve starts from
$L_{q-v}=0$ and ends at $L_{\rm max}$, 
then it comes back toward the small $L_{q-v}$ side along
the second curve with larger $E$ and it stops at $L_{q-v}=L_1(>0)$. 
Then it turns to the larger $L_{q-v}$ again along the third curve with 
more large $E$, and
it stops at $L_{q-v}=L_2(<L_{\rm max})$. (See Fig. \ref{EL-005all}.)
This behavior is common to all
the baryon configurations of different $k'$s, and the slopes of the curves 
of different $k$s are proportional to $k$. 

However we find a real screening point at $L^*(\ll L_{\rm max})$, where
the baryon decays to the quarks and gluons,
before arriving at $L_{\rm max}$. So we could not see the above complicated
properties as stable baryon states, and
for $L^*<L_{q-v}$, they decay into the quarks and gluons. When T increases and
approaches to $T_{\rm melt}$, then $L^*$ vanishes to zero and all the baryons
melt down.

In the section \ref{finite-t} we review our model briefly,
and the equation of motion for D5 branes is solved to show four types
of solutions and to discuss the stability of the baryon. 
In the section 3, the balance condition of the forces is given, and in the
next section 4, $E-L_{q-v}$ relations and the estimation of the screening mass
are given. In the section 5, we discuss the stability of the baryon state
from the viewpoint of baryon vertex energy and the gauge condensate parameter.
The summary and discussions are given in the final section.

\section{D5 baryon vertex at finite temperature}\label{finite-t}
Here, we consider the baryon configurations in the non-confining, finite temperature
Yang-Mills theory.
Such a model is given by
the AdS blackhole solution, which represents the high temperature
gauge theory. In our theory with dilaton, the corresponding
background solution is given as
\cite{GSUY}
\beq
ds^2_{10}= e^{\Phi/2}
\left(
\frac{r^2}{R^2}\left[-f^2dt^2+(dx^i)^2\right] +
\frac{R^2}{r^2f^2} dr^2+R^2 d\Omega_5^2 \right) \ .
\label{finite-T}
\eeq

\beq 
f=\sqrt{1-({r_T\over r})^4}, \quad e^\Phi= 
1+\frac{q}{r_T^4}\log({1\over f^2}) \ , \quad \chi=-e^{-\Phi}+\chi_0 \ ,
\label{finite -T-dilaton} 
\eeq
The temperature ($T$) is denoted by $T=r_T/(\pi R^2)$. The world volume 
action of D5 brane is rewritten by eliminating the $U(1)$ flux in terms
of its equation of motion as above, then we get its energy as \cite{GI}
\be \label{u-T}
U_{D5} = {N\over 3\pi^2\alpha'}\int d\theta~e^{\Phi/2} f
\sqrt{r^2+r^{\prime 2}/f^2 +(r/R)^{4}x^{\prime 2}}\,
\sqrt{V_{\nu}(\theta)}~.
\ee
where 
\be\label{PotentialV}
V_{\nu}(\theta)=D(\nu,\theta)^2+\sin^8\theta
\ee
\be \label{d}
D\equiv D(\nu,\theta) = \left[{3\over 2}(\nu\pi-\theta)
  +{3\over 2}\sin\theta\cos\theta+\sin^{3}\theta\cos\theta\right].
\ee
Here, $D$ denotes the electric flux at $\theta$-cross section. And
the integration constant $\nu$ is expressed as $0\leq\nu=n/N\leq 1$,
where $n$ 
corresponds to the number of F-strings emerging from one of the pole of
the ${\bf S}^{5}$ at $\theta=0$. Then $N-n$ comes out from the other
pole at $\theta=\pi$.

\vspace{.3cm}
\subsection{Embedded solutions of the D5 brane}\label{D5-embedding}
The vertex configuration is obtained from the above Legendre transformed action
(\ref{u-T}).
Here for simplicity,
we concentrate on the point vertex configuration and $\nu=0$, which means
the quarks come from the cusp at $\theta=\pi$. There is no other cusp.
Then, we set as $x'=0$, and we obtain
\be \label{u-T-2}
U_{D5} = {N\over 3\pi^2\alpha'}\int d\theta~e^{\Phi/2} f
\sqrt{r^2+r^{\prime 2}/f^2}\,
\sqrt{V_{0}(\theta)}~.
\ee
 From this action,
the equation of motion of $r(\theta)$ is obtained as follows
\beq\label{r-eq-T}
\partial_{\theta}\left({r^{\prime}\over\sqrt{r^2f^2+(r^{\prime})^2}}\,
\sqrt{V_{0}(\theta)}\right)
-
{P(r)\over\sqrt{r^2f^2+(r^{\prime})^2}}\,
\sqrt{V_{0}(\theta)}=0~.
\eeq
\beq\label{g-func}
P(r)={1\over 2e^\Phi}\partial_r\left(e^\Phi ~r^2f^2\right)
\eeq

\begin{figure}[htb]
\centerline{{\epsfxsize=8cm\epsfbox{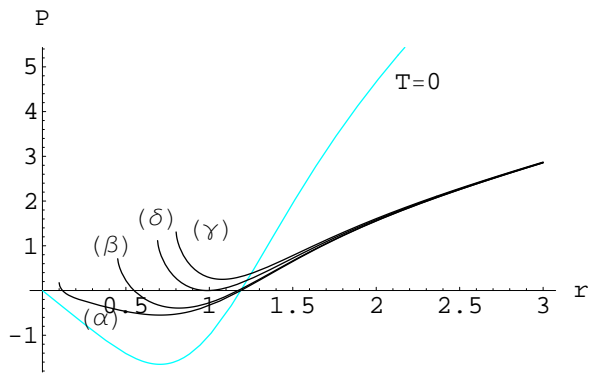}}
}
\caption{$P(r)$ for $q=2$ and $r_T=(\alpha)~ 0.1,~(\beta)~ 0.45,~(\gamma)~ 0.8,
~(\delta)~0.689 $. The curve $(\delta)$ is given at the critical
point $T_{c_1}$. The curve of $T=0$
shows $3P\vert_{r_t=0}=3r(1-q/r^4)/(1+q/r^4)$.} 
\label{p-r}
\end{figure}

The solutions $r(\theta)$ of equation (\ref{r-eq-T}) are characterized
by the prefactor $P(r)$ in the second term. Its behavior depends on the 
temperature. For the comparison, $P(r)$ at T=0 is also shown in the
Fig.~\ref{p-r} with the one of the various finite temperatures. 

\begin{figure}[htb]
\centerline{{\epsfxsize=7cm\epsfbox{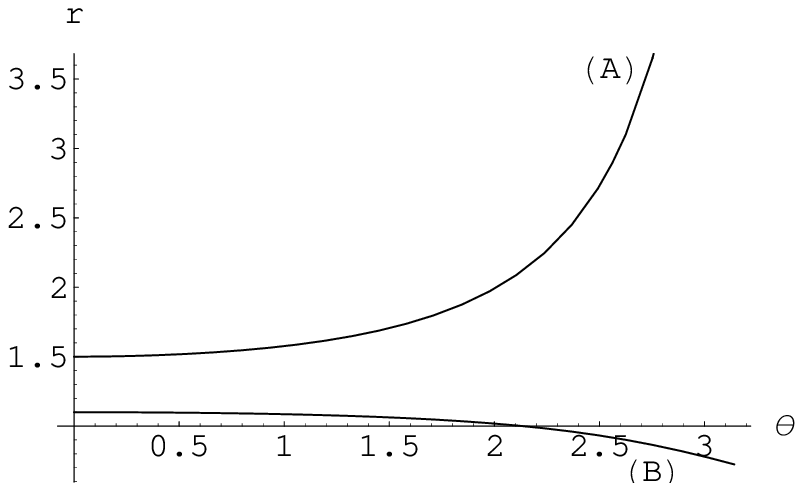}}
{\epsfxsize=7cm\epsfbox{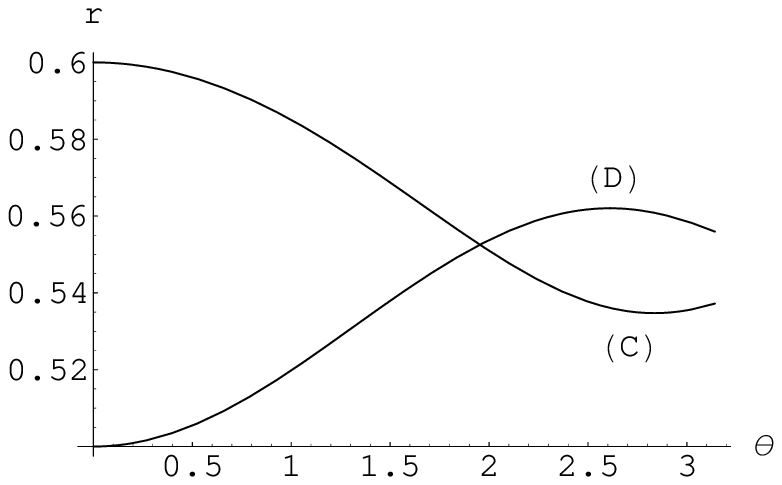}}
}
\caption{The four types of solution of
$r(\theta)$ for $q=2$ at $r_T=0.45$. The curves represent for  
(A) $r_0=1.5$, (B) $r_0=1.1$, (C) $r_0=0.6$ and (D) $r_0=0.5$  respectively.} 
\label{sol-r1}
\end{figure}

At any temperature,
$P(r)$ is positive at large r and changes its sign to the 
negative value for the low temperature cases according to the curves
$(\alpha)$ and $(\beta)$ in the Fig.~\ref{p-r}. When we take the value of
$r(0)(\equiv r_0)$ at the point of $P>0$ ($P<0$), then  
we obtain the typical
configuration (A) ((B)) of the baryon as shown in the Fig.~\ref{typical-sol-1} where
F-strings are added according to the condition given below. Here we notice
only the configuration of the vertex part. For the case of zero temperature,
the two types solutions (A) and (B) are enough correspondingly to
the region of $P>0$ and $P<0$ respectively.
However, at finite and low temperature, the behavior of $P$ is not so simple
since the zero point appears at smaller $r$. When $r_0$ is chosen
near this zero point, we find two typical solutions (C) and (D) for 
$P<0$ and $P>0$ respectively as shown in the Fig.~\ref{sol-r1}. They oscillate
once with respect to $\theta$, and these solutions can not be seen at zero 
temperature.

\begin{figure}[htb]
\centerline{
{\epsfxsize=14cm
\epsfbox{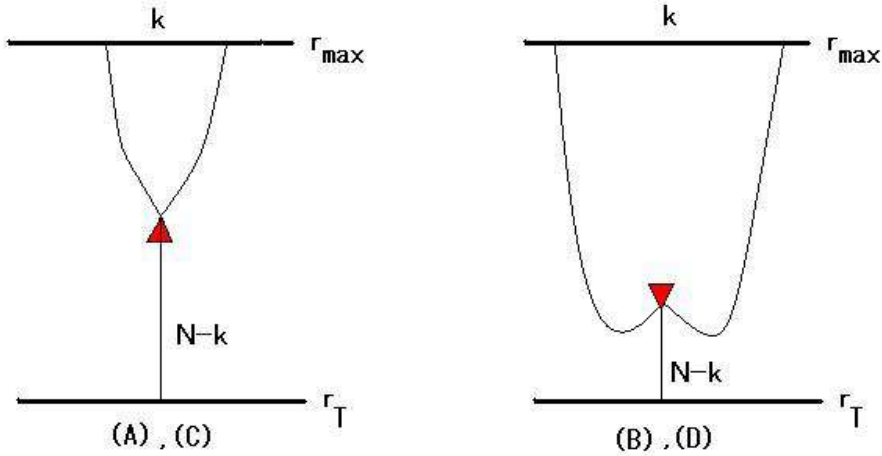}
}}
\caption{\small $k$-quarks baryons at finite temperature. The triangles represent
the end point $r(\pi)$ of D5 vertex 
and the direction of the cusp at $\theta=\pi$. The left (right) 
corresponds to the solutions
(A) and (C) ((B) and (D)).} 
\label{typical-sol-1}
\end{figure}

\begin{figure}[htb]
\centerline{{\epsfxsize=8cm\epsfbox{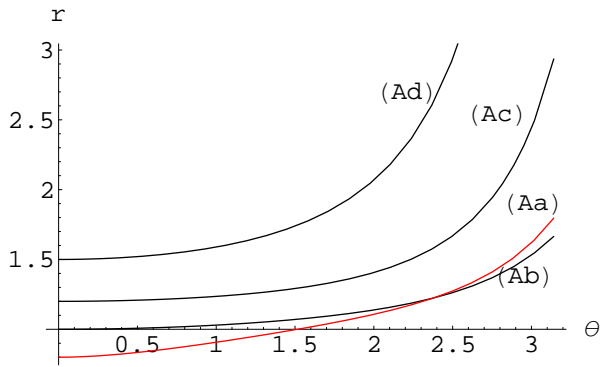}}
}
\caption{The type (A) solutions of
$r(\theta)$ for $q=2$, $r_T=0.8$ and (Aa) $r_0=0.80001$ (Ab) $r_0=1.0$
(Ac) $r_0=1.2$ (Ad) $r_0=1.5$.} 
\label{sol-r4}
\end{figure}

At higher temperature, $T>T_{c_1}$,
the zero points of $P(r)$ disappear as shown by
the curve $(\gamma)$ in the Fig.~\ref{p-r}.
In this case, only the type (A) solutions are obtained, and several solutions
with different $r_0$ 
are shown in the Fig.~\ref{sol-r4}. We can see that the end point value 
$r(\pi)\equiv r_c$
of the solutions $r(\theta)$
takes its minimum for some initial value
$r(0)$, then it rises when $r(\pi)$ decreases furthermore and
approaches to the horizon $r_T$. 

\vspace{.5cm}

\vspace{.3cm}
Next, we turn to the F-strings.
In the present
finite $T$ model, the effective quark mass $(\tilde{m}_q)$ is finite
and it diverges at $T=0$. In this sense,
the present theory is in the quark deconfinement phase for finite $T$. So
free quarks or independent F-strings could exist, and the end points of these
F-strings touch on the horizon $r_T$ and $r_{\rm max}$. 
Then, some of the F-strings, which are originally attached to the baryon, could 
escape from the baryon to form such free F-strings. And
as shown in the Fig.~\ref{typical-sol-1}, 
we could find $k(<N)$-quarks baryon
configuration with extra $N-k$ F-strings which connect the vertex and
the horizon. These $N-k$ F-strings are not observed as quarks in the gauge theory,
but they contribute to the energy of the baryon.
As a result, some number of $N-k$ quark disappear from the original baryon, 
and we would find it as a $k$-quarks baryon. Then,
this implies that we may find color non-singlet baryons as
depicted in the Fig.~\ref{typical-sol-1}.



\vspace{.5cm}
\section{No force condition and $k$-quarks baryon}
{Full baryon configuration} is given by the vertex and F-strings.
The configurations are determined through the no force condition
at the cusp point of the vertex and the profiles of the vertex and
the F-strings. They are characterized by the energy of the total system,
the baryon energy, and the distance between the quark and the vertex ($L_{q-v}$).

\vspace{.5cm}
The tension of D5-brane at
$r_c$ is given as
\begin{equation}
\frac{\partial U_{D5}}{\partial
  r_c}=NT_Fe^{\Phi/2}\frac{r'_c}{\sqrt{r'^2_c+r^2_cf(r_c)^2}}
\end{equation}

As for the F-string  The  energy  is written as
\begin{equation}
U_{F}=kU_{F}^{(1)}+(N-k)U_F^{(2)}
\end{equation}

Here, $U_{F}^{(1)}$ is the energy of string  which extends to the
$x$-direction  and 
its  action is 
written as 
\begin{equation}
S_{F}^{(1)}=-\frac{1}{2\pi\alpha'}\int dtdx
e^{\Phi/2}\sqrt{r^2_x+f^2(r)(r/R)^4}
\end{equation}
where $r_x=\frac{\partial r}{\partial x}$ and the world sheet
coordinates are set as $(t,x)$. Then, its energy and $r$-directed
tension at the point $r=r_c$ are obtained as follows
\begin{equation}
U_F^{(1)} =T_{F}\int dx e^{\Phi/2}\sqrt{r^2_x+f^2(r)(r/R)^4} \label{uf}
\end{equation}
\begin{equation}
\frac{\partial U_F^{(1)}}{\partial r_c} =T_F\frac{r_x}{\sqrt{r^2_x+(r_c/R)^4f(r_c)^2}}\end{equation}

$U_F^{(2)}$ is the energy of string extending  from horizon $r_T$ to
$r_c$, and it is written as
\begin{equation}
U_{F}^{(2)}=T_F\int_{r_T}^{r_c} dr e^{\Phi/2} 
\end{equation}

Then the no force condition is obtained as
\begin{equation}
N\frac{r'_c}{\sqrt{r'^2_c+r^2_cf(r_c)^2}}+(N-k)=k\frac{r_x}{\sqrt{r^2_x+(r_c/R)^4f(r_c)^2}} \label{no-force}
\end{equation}

\subsection{Lower bound on $k$} \label{bound-k}
 From this equation, we can estimate the lower bound for the k, which is
written as 
\beq
 k\geq {N\over 2}(1+Q_5)\, , \quad
 Q_5\equiv \frac{r'_c}{\sqrt{r'^2_c+r^2_cf(r_c)^2}}\, .
\eeq

The value of $Q_5$ depends on the parameters, but we can see it does not
arrive at -1. Then $k>0$, however, $k$ could decrease and approaches to zero
by choosing the parameters. For example, we find $Q_5=-.99916$ for 
$(q,r_0,r_T)=(2.0,~0.6,~0.01)$. 

In our model, the F-string configuration must satisfies Eq.(\ref{const})
with a constant $h$. This implies $h=0$ for the string which touches on the horizon 
$r_T$ since $f(r_T)=0$. So we obtain a vertical straight line as the configuration.
However, it would be possible to obtain a slightly curved string configuration in a
more realistic model. Then the configuration of $k=0$ would be possible 
in principle at finite temperature. But it would not realized energetically
as a stable state.


\subsection{Energy of $k$-quarks baryon}
The energy of baryons are constructed by the F-strings energy and
that of the vertex. 
In obtaining the F-string configuration from Eq.
(\ref{uf}), we introduce the constant $h$ as follows,
\begin{equation}\label{F-stringh}
e^{\Phi/2}\frac{r^4f^2(r)}{R^4\sqrt{r^2_x+(r/R)^4f^2(r)}}=h \label{const}
\end{equation}

\vspace{.3cm}
For $r_x>0 (<0)$, $k$-quarks baryons configuration becomes the type (A)((B)).
First, consider the solution of (A).  In this case, 
the energy of the F-strings is given as
\begin{equation}
U_{F}=T_F\left[k\int_{r_c}^{r_{max}}dr\frac{e^{\Phi/2}}{\sqrt{1-\frac{h^2R^4}{f^2e^{\Phi}r^4}}}+(N-k)\int_{r_T}^{r_{c}}dre^{\Phi/2}\right] 
\end{equation}
The second term denotes the F-strings stretching between horizon and $r_{c}$.
And the distance $L_{q-v}$ is given as
\begin{equation}
 L_{q-v}=R^2\int_{r_{c}}^{r_{max}}dr\frac{1}{r^2f^2\sqrt{\frac{e^{\Phi}r^4f^2}{h^2R^2}-1}}
\end{equation}

\vspace{.3cm}
Next we turn to the configuration of (B). 
In this case, the string configuration includes a bottom part of U-shaped
configuration and the bottom point is seen at $r=r_{min}$, where we find
\begin{equation}
\left.\frac{dr}{dx}\right\vert_{r=r_{min}}=0
\end{equation}
Since (\ref{F-stringh}) can be used at any $r$, we obtain
\begin{equation}
h=e^{\Phi(r_{min})/2}f(r_{min})(\frac{r_{min}}{R})^2
=e^{\Phi(r_c)/2}\frac{r_c^4f^2(r_c)}{R^4\sqrt{r_x^2+(r_c/R)^4f^2(r_c)}}
\end{equation}
 From this, $r_{min}$ is determined by
$r_c$, then by $r_0$. On the other hand,
$r_x$ is determined by using $r_c$ from ($\ref{no-force}$).
As a result, we obtain the string energy and the distance 
between the quark and the vertex as follows,
\begin{equation}
U_F=T_F\left[k\left(\int_{r_{min}}^{r_{max}}dr\frac{e^{\Phi/2}}{\sqrt{1-\frac{h^2R^4}{f^2e^{\Phi}r^4}}}+\int_{r_{min}}^{r_c}dr\frac{e^{\Phi/2}}{\sqrt{1-\frac{h^2R^4}{f^2e^{\Phi}r^4}}}\right)+(N-k)\int_{r_T}^{r_c}dr
e^{\Phi/2}\right]
\label{efb}
\end{equation}
\begin{equation}
L_{q-v}=R^2\left[\int_{r_{min}}^{r_{max}}dr\frac{1}{r^2f\sqrt{\frac{e^{\Phi}r^4f^2}{h^2R^4}-1}}+\int_{r_{min}}^{r_{c}}dr\frac{1}{r^2f\sqrt{\frac{e^{\Phi}r^4f^2}{h^2R^4}-1}}\right]
\end{equation}
Here we consider the energy of $k$-quark (for any $k$)
baryon and $N-k$ free quarks to
compare the energy with that of the $N$ free quarks
which form the baryon as a singlet state.
So we add $(N-k)$
free quarks, which are strings stretching between horizon and $r_{max}$,
to the $k$-quark baryon state.
We estimate the energy of the baryon in this way as a
function of the distance $L_{q-v}$.
Then the total $k$-quark baryon energy is obtained  as 
\begin{equation}
E_k=U_{F}+U_{D5}+(N-k)\tilde{m}_q\, ,
\label{totu}
\end{equation}
where notice that the first two terms $U_{F}$ and $U_{D5}$ are dependent
on k, and $\tilde{m}_q$ is the energy of a free quark,
\begin{equation}
\tilde{m}_q=T_F\int_{r_T}^{r_{max}} dre^{\Phi/2}\, .
\end{equation}

\section{Color screening and baryon melting}
Here we consider the stability of the various baryon configurations
in the deconfining thermal medium. Although the baryon is in the deconfining 
phase some of the baryon states could be stable
until the temperature arrives at the critical point where
all the baryons melt down to free quarks. This melting temperature is
estimated below by comparing the baryon energy and the energy of free
quarks.

In the followings, we estimate numerically 
the energy $E$ and the distance between quark
and the vertex, $L_{q-v}$ for $k$-quark baryon ($k\leq N$) 
in order to cover various cases. 
The color screening due to the temperature is observed as
the existence of a maximum value of $L_{q-v}$ for each baryon states.
However, the real maximum value of $L_{q-v}$ is determined energetically
by comparing the energy of the baryon state and the one of the
summation of the free quarks.

\subsection{For $k=N$ case}
\begin{figure}[htb]
\centerline{{\epsfxsize=7cm\epsfbox{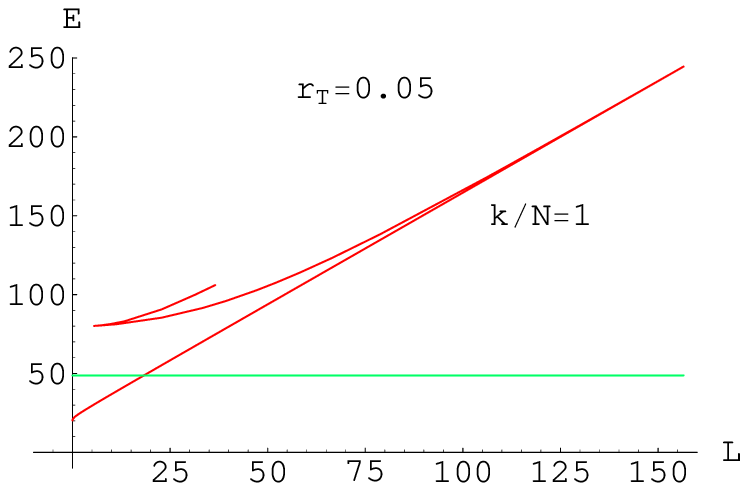}}
{\epsfxsize=7cm
\epsfbox{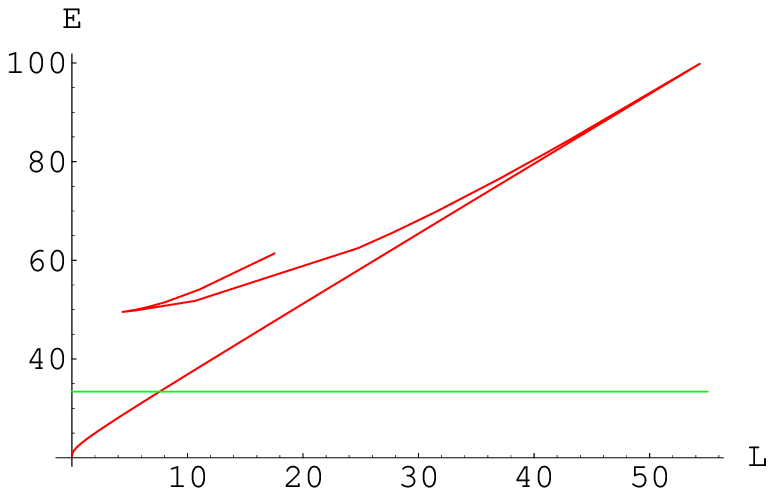}
}}
\caption{Energy of $N$-quarks baryon versus $L_{q-v}$ (which is
denoted simply by L in the figure) at $r_T=0.05$ (left)
and $r_T=0.1$ (right) for $q$=2. 
In these figures two turnovers are seen as mentioned in the text. 
However, physically allowed region is only the part below the horizontal 
lines of $N\tilde{m}_q$. 
} 
\label{EL-005all}
\end{figure}

Typiclal $E-L_{q-v}$ curves are shown in the Fig.~\ref{EL-005all}
for two temperatures. The beginning point at $L=0$ corresponds to 
the solution of $r(\theta)$ with $(r(\pi)=)r_c=r_{\rm max}$ and
$(r(0)=)r_0>r_T$. Then $E$ increases almost linearly with $L_{q-v}$. 
We notice two
turn over points in these curves, and the second one has not been seen
in the case of the mesons. We could see that the second turn over 
is related to the existence of two zero points
of $P(r)$ and new two type of solutions ((C) and (D))
as seen in the section \ref{finite-t}. 
However, this intriguing
behavior of $E-L_{q-v}$ curves
is seen in the unstable region of the baryon since the baryon energy
in this region is higher than the sum of the free quarks energy, which are shown by
the horizontal lines. They cross at $L=L^*$, which depends on the temperature,
then the baryon, which has the energy
below $N\tilde{m}_q$, is allowed as a stable state. 
The point $L^*$ is therefore interpreted as
the maximum size of the baryon. Within this distance, the color force
works on the quark, and the color gauge force is screened outside this distance. 
As a result, all the baryons decay
into quarks and gluons for $L^*<L_{q-v}$. 

\begin{figure}[htb]
\centerline{{\epsfxsize=7cm\epsfbox{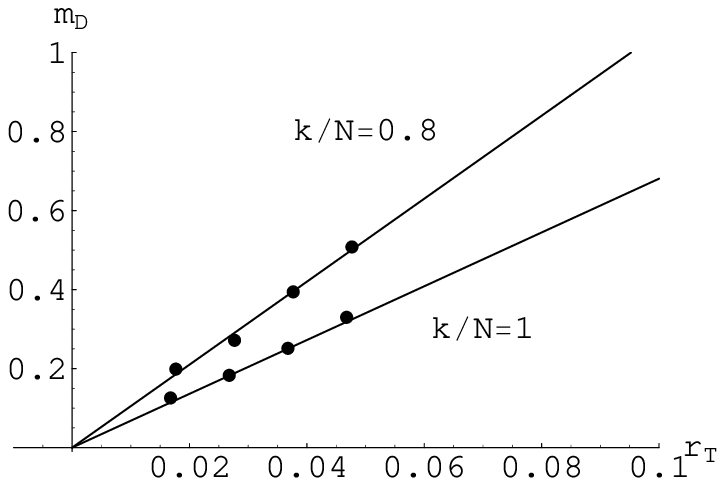}}
{\epsfxsize=7cm\epsfbox{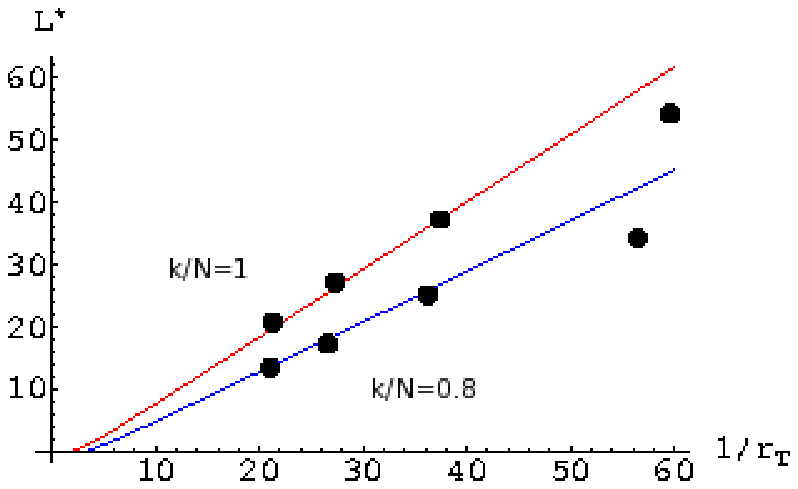}}
}
\caption{The left hand side shows the Debye screening mass $m_{D}$ estimated for 
$k/N=1$ and 0.8. 
And the line is fitted as $m_D=6.81 r_T$ ($m_D= 10.5~r_T$)
for $k/N=1$ ($k/N=0.8$). In the right hand figure,
melting point $L^*$ are shown by the curves
for $k/N=1$ and 0.8 baryon versus $1/r_T$ 
at $q=2$. The dots represent the superposition of the data for
$m_D$ given in the left hand figure. They are given
as $L^*=6.8/m_{D}$ for both $k/N=1$ and 0.8.
}
\label{rt-Lmax}
\end{figure}

Of course, the value of $L^*$ depends on the temperature. Their values
for various temperatures are plotted in the Fig.~\ref{rt-Lmax}.
The numerical results show 
that $L^*$ is proportional to $1/r_T$ or $1/T$, and it approaches to zero at
some limiting temperature. Above this temperature,
any baryon state is not allowed.

The value of $L^*$ should be related 
to the Debye screening. 
In the gauge theory,
the Debye screening has been considered 
as the correction to the gluon propagator. Then the potential between two 
colored particles separating $r$ is given by the form $e^{-m_{D}r}/r$.  
However, in the present model, the force at short distance is given by
the strong coupling CFT and a long range force which leads to linear potential.
And we find a linear-like potential near $L^*$ 
even if it is still not so large. 
So the potential screened by the thermal effect is just this linear 
like potential, which is considered as the tail of the 
linear potential seen at large $L_{q-v}$ in the case of $T=0$.
When we denote this tail as $V_0(L)$ at zero temperature,
it is modified at finite temperature as,
\beq
 V\equiv E-N\tilde{m}_q=V_0(L)~e^{-m_D(T) L}\, ,
\eeq 
where $m_D(T)$ denotes the Debye mass. 
Using this formula, the numerical curves of $E-L$ are fitted by this form,
and we obtain $m_D(T)$. However we used here $V_{T_{\rm min}}(L)$ instead of
$V_0(L)$, where $V_{T_{\rm min}}(L)$ denotes the potential of the lowest
temperature. So we corrected the data of $m_D(T)$ obtained by shifting
the temperature $T\to T'$ such that we could obtain $m_D(T')\to 0$ for
$T'\to 0$. The results are shown in the left hand of the Fig.\ref{rt-Lmax}.

Then we could find the relation $L^*=c/m_D(T)$ with a definite 
constant $c=6.8$.  And we obtain the result that $m_R(T)$ increases 
linearly with the temperature T.

\vspace{.3cm}
In \cite{BKY}, using the AdS/CFT correspondence, Bak, Karch and Yaffe examined 
the behavior of correlators of Polyakov loops and other operators in 
${\cal N}=4$ supersymmetic Yang-Mills theory at non-zero temperature. 
And the implications for Debye screening in that strongly coupled non-Abelian 
plasma, and comparisons with available results for thermal QCD, were 
discussed.  The connection with our result is that the proportionality 
$m_D\propto T$ is common with \cite{BKY}.  In \cite{AS}, however, $m_D/T$ 
is predicted to decrease slowly with $T/T_c$ on the basis of nonperturbative 
computation in the deconfined phase of QCD, using the method called background 
perturbation theory.  This prediction is in good agreement with the data of 
lattice QCD \cite{KZ}.  In order to reproduce this decrease, string-string 
interactions are taken into of, in string side.  This is a future problem.   

\subsection{For $k<N$ case}

\begin{figure}[htb]
\centerline{
{\epsfxsize=7cm\epsfbox{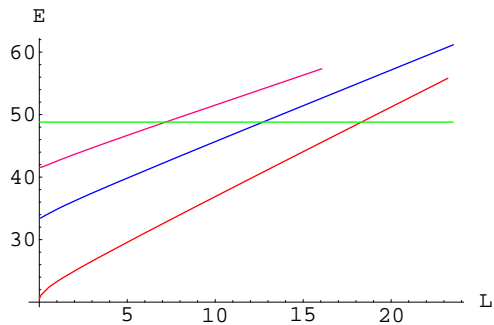}}
}
\caption{Energy of $k$-quarks baryon versus $L_{q-v}$ for $r_T=0.05$.
The curves represents $k/N=1$ (a), $0.8$
(b) and $0.67$ (c). And the flat line represents $N\tilde{m}_q=48.8066$.
} 
\label{EL-LT}
\end{figure}

In this subsection, we discuss the properties discussed above for 
the baryon with $k<N$. Several $E-L_{q-v}$ curves for different $k$-baryons
are shown in the Fig.~\ref{EL-LT}.
The melting points $L^*$ are obtained from these results, and
they are shown for $k/N=0.8$ in the Fig.~\ref{rt-Lmax}. Compared to the case
of $k/N=1$, it starts from a larger $1/r_T$ (smaller temperature) with a
smaller grade in the $L^*$--$1/r_T$ plane. For smaller $k$, the 
starting point moves to larger $1/r_T$ with smaller slope. Then we could find
small $k$-baryon at low temperature as stated in the subsection \ref{bound-k}.

However, the energy of $k$-baryon is larger than the $k'$-baryon for
$k<k'$ at the same $L$. Then the $k$-baryon ($k<N$) is always 
observed as an exited state
of the $N$-quark baryon. 
One more point to be noticed is that our model
is in the deconfinement phase as far as the temperature is finite. However,
the confinement phase would be realized at a finite temperature, 
in other words, the critical temperature of quark confinement and 
deconfinement must be finite. In this sense, the lower bound of $k$ would be
existing at some finite value. But this bound is not given here.  

\section{Baryon stability}

We discuss here more about the stability of the baryon from two viewpoints
in order to get physical insights. Since the vertex is an important part of the 
baryon, it would be important to see its role for the baryon stability. Another
important ingredient in our model is the gauge condensate which is the main
source of the confinement force. So we like to see its role in the deconfinement
phase considered here.

\subsection{Vertex energy and stability}
First, we give a comment from the viewpoint of the vertex energy.
In \cite{GSUY}, a comment on the stability of the baryon at finite temperature 
has been given through the vertex energy without any inner structure.
We give a similar kind of comment on this point in terms of the energy of the vertex
with structure given above. Before doing it, we firstly notice that the energy
of the vertex without structure. Assuming the baryon is at $r$, the energy
is given from 
equation (\ref{u-T-2}) by setting $r'=0$ as,
\beq\label{v-energy}
 U_{D5}=\tilde{N} e^{\Phi/2} r f(r)
\eeq
where $\tilde{N}={N\over 3\pi^2\alpha'}\int d\theta~
\sqrt{V_{0}(\theta)}$ denotes a $r$-independent constant. Then we find
\beq
 P(r)\propto {\partial U_{D5}\over \partial r}.
\eeq
 From this, we can say that the zero point of $P(r)$ is needed to find a
minimum of $U_{D5}$, and we actually find such a point at low energy.
However, there is a temperature $T_{c_1}$ above which 
the zero point disappears, then the vertex becomes unstable as mentioned
in \cite{GSUY}. However, this
statement should be modified when the structure of the vertex is taken
into account of.  

\begin{figure}[htb]
\centerline{{\epsfxsize=8cm\epsfbox{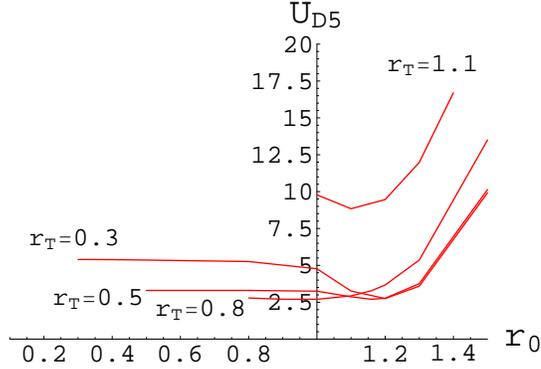}}}
\caption{The vertex energy $U_{D5}$ versus $r_0$. The minimum of $U_{D5}$
exists near $r_0=q^{1/4}$. For $r_T\geq 0.8$, the minimum increases rapidly.} 
\label{U-D5-r0}
\end{figure}

By substituting the solution of (\ref{r-eq-T}) into (\ref{u-T}), we obtain
the vertex energy $U_{D5}$ as a function of 
$r_0$. In this case, we find the minimum even if the temperature
is taken to be larger than $T_{c_1}$, 
as seen in the Fig.~\ref{U-D5-r0}. Then we need another criterion to see
the stability of the baryon. 

\begin{figure}[htb]
\centerline{{\epsfxsize=8cm\epsfbox{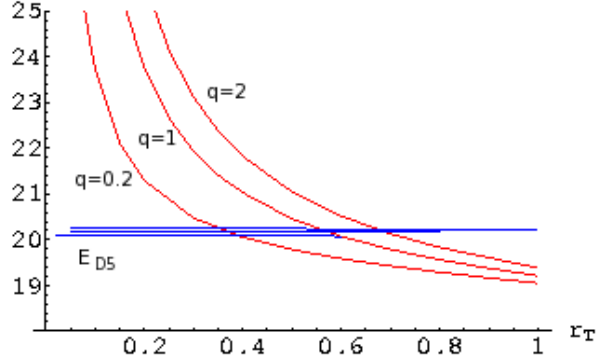}}
}
\caption{The T dependence of the 
minimum energy of the $N$-quark baryon ($E_{D5}$) are shown by the top, 
middle and the bottom lines for $q=2.0,~1.0$ and $0.2$ respectively.
The three curves show $N\tilde{m}_q$ for $q=2.0,~1.0$ and $0.2$
} 
\label{ED5-mq}
\end{figure}

The new criterion can be stated as follows. The stability of the baryon
is assured when its energy is smaller than the sum of free effective-quark mass,
$N\tilde{m}_q$. Because the baryon decays to
free quarks when its energy exceeds $N\tilde{m}_q$. 

The energy of the baryon is given 
by the sum of $E_{D5}$ and F-strings. Then the minimum value of the baryon
energy is given by $E_{D5}^{\rm min}$ when F-strings part vanishes due to $L=0$.
Then, the stability of the baryon is found by comparing $E_{D5}^{\rm min}$ 
with $N\tilde{m}_q$. 
The value of $E_{D5}^{\rm min}$
is given by the D5 energy with a configuration of $r_T<r(o)<r<r(\pi)=r_{\rm max}$.
In this configuration, the F-strings are reduced to a point
at $r_c$.
The higher energy states are obtained by adding
F-strings with finite length to the D5
brane with $r(\pi)<r_{\rm max}$ and $r(0)\geq r_T$.

In the Fig.~\ref{ED5-mq}, the 
minimum energy of the $N$-quark baryon ($E_{D5}$) and $N\tilde{m}_q$
are plotted as functions of 
$r_T$. We find the melting point ($T_{\rm melt}$)
near $T_{c_1}$, above which $E_{D5}$
exceeds $N\tilde{m}_q$. Then the baryon decays to quarks and gluons. 
It is an interesting point that this temperature is very near to $T_{c_1}$.
This temperature
$T_{\rm melt}$ depends sensitively
on the gauge condensate $q$, so $q$ is also related
to the stability of the baryon state. On this point, we discuss in the
following subsection.

\subsection{Gauge condensate and Baryon at high temperature}

As shown in the Fig.~\ref{ED5-mq}, $T_{\rm melt}$ increases with 
$q$. Then an unstable baryon at a temperature can be stabilized by increasing
the gauge condensate $q$. This point is understood in terms of the
role of the gauge condensate in the gauge theory. 

The role of $q$ has been made clear as a source of 
quark confinement force. At zero temperature with finite $q$, we find the linear
potential between quark and anti-quark in the mesons \cite{GY}
and also between the vertex
and the quark in the baryons \cite{GI}. The tension of this linear potential
is proportional to $q^{1/2}$ in both cases. On the other hand, the role of
the temperature is to screen the color force and to prevent making the bound
states of quarks. The highly exited state has a large 
size, then it would be destabilized by the screening effect at high
temperature. However, this temperature would depend on the magnitude of the force
or the value of $q$.

Actually, we find that the strength of the linear potential 
preserved even in the short range is proportional to $q^{1/2}$, so the force to
make the baryon at finite T is also proportional to $q^{1/2}$. Then we need
higher temperature to destroy the baryon formed at larger $q$.
Since the effects of $T$ and $q$ are opposite, then the melting point 
$T_{\rm melt}$ increases with $q$.

This implies that the baryon 
configuration constructed by the vertex and F-strings can not be obtained 
for $q=0$. In other words, the gauge condensate is essential to find
k$(\leq N)$-quark baryon at finite temperature. 
On the other hand, mesons are expected to be observed even in the deconfinement 
high temperature phase of $q=0$ \cite{GSUY}.

This is understood also from the vertex energy given by (\ref{v-energy}) which
is represented for $q=0$ as follows,
\beq\label{v-energy-2}
 U_{D5}=\tilde{N} \sqrt{r^2-{r_T^4\over r^2}}\, .
\eeq
This implies that the vertex is not stable and it vanishes into the horizon.
But this result is given for the structure less vertex \cite{GSUY}. When the
structure is considered, the end point of the vertex increases
infinitely toward the boundary when we solve the configuration of
the D5 brane, $r(\theta)$. The reason is that the mass scale to retain $r_c$
at a finite value is given by $q$. For $q=0$, the configurations given above
do not exist. In other words, the confinement force given at zero temperature
should be retained at finite temperature in order to get k$(\leq N)$-quark baryon.
This point should be verified by the experiments at the finite temperature.


\section{Summary and Discussion}
The baryon is studied here from a holographic approach.
It has a complicated structure since it is constructed by 
the $N$-quarks and their vertex, which would not be
a point but an extended object. From the holographic viewpoint, the baryon vertex
is identified with the D5 brane and its configuration is obtained by
solving the equation of motion given by D5 brane action which includes 
N fundamental string flux. At zero temperature,
this vertex has two typical structures. When we observe it in our real
space-time, its shape is a point like or an one dimensionally extended object. From
10D supergravity side, both configurations have further complicated structure
due to the extra six dimensional coordinates.

For the simplicity, here, we study the baryon with the vertex, which appears as 
a point in the 4D space-time, at finite temperature. In our 
model, the quarks are not confined in the sense that the effective free quark 
mass is finite. In this sense, the theory is in the 
deconfinement phase. Then, the fundamental strings could touch on the
horizon, then the baryon configuration with $k(<N)$-quark is possible in this phase.
The configuration of the $k$-quark baryon is formed by the D5 brane vertex
and k fundamental strings which connect the cusp of the vertex and the boundary.
The boundary is here set at the finite radial coordinate of the extra dimension
as a cutoff point of the string energy or the position of the flavor quark brane.
And the remaining $N-k$ fundamental strings are connecting the vertex and
the horizon.
Then the $k$-quark baryons with $k\leq N$ in a thermal medium are studied here.

The structure of the vertex becomes a little rich at finite temperature
since two extra new solutions of vertex configuration are found. They
are not seen at zero temperature. However, we find that
these new configurations can not
be observed since they are unstable against to the decay
into the free quarks. This instability is studied through
the calculation of the baryon energy for its fixed size $L$. The size is defined
by the distance between the vertex and a quark. We assumed that
the quarks in the baryon are distributed
maximally symmetric, then we impose the no force condition 
given at the connecting point of the fundamental strings and the vertex. 
The energy of $k$-quark baryon is obtained in this way and it is compared with the
one of the free $k$-quarks. 

 From these analyses, the baryon energy is smaller than the one of the
free quarks at small $L$. And
we find a point $L=L^*$, where the baryon energy exceeds the one of the
free quarks. Then the baryon with large size $(L^*<)L$ decays
into the free quarks, we call this point as melting point. The size depends
on the vertex configuration through the no force condition, and the configurations 
newly found at finite temperature give the larger size than $L^*$.

The value of $L^*$ can be also
related to the Debye length, which has been introduced originally for the Coulomb
potential in the electro-magnetism, and it depends on the temperature. 
This is estimated here by multiplying the Gauss damping factor to the linear-like
potential, which appear as the tail of the zero temperature linear potential. And
we find the expected behavior of the Debye mass gap, which is proportional to
the temperature. This temperature dependence is seen for all the state of
$k$-quark baryon, but small $k$-quark baryons are restricted to low temperature.

Finally, we find that the baryon stability at finite temperature
is supported by the gauge condensate $q$.
The thermal medium destabilizes the baryon by the 
screening effect of the color force, but $q$ support the color force
to make a linear potential to retain the baryon at not so high temperature,
and we could observe the baryon even at finite temperature.

\section*{Acknowledgements}
We thanks to M. Tachibana for useful discussions and comments. 
The work of M. Ishihara is supported by JSPS Grant-in-Aid for Scientific
Research No. 20 $\cdot$ 04335.

\vspace{.5cm}

\end{document}